\documentclass[prl,twocolumn,superscriptaddress,nofootinbib,longbibliography]{revtex4-1}

\usepackage{graphicx,epsfig}
\usepackage{color}
\usepackage[usenames,dvipsnames]{xcolor}
\usepackage{amsmath,amssymb, amsthm}

\usepackage{numprint}

\def\bb{\begin{eqnarray}}
\def\ee{\end{eqnarray}}
\newcommand{\ket}[1]{| #1 \rangle}
\newcommand{\bra}[1]{\langle #1 |}

\begin{document}

\title{Quantifying the quantum heat contribution from a driven superconducting circuit}

\author{Cyril Elouard}
\email{cyril.elouard@gmail.com}
\affiliation{Department of Physics and Astronomy, University of Rochester, Rochester, NY 14627, USA}
\author{George Thomas}
\affiliation{QTF centre of excellence, Department of Applied Physics, Aalto University School of Science, P.O. Box 13500, 00076 Aalto, Finland}
\author{Olivier Maillet}
\affiliation{QTF centre of excellence, Department of Applied Physics, Aalto University School of Science, P.O. Box 13500, 00076 Aalto, Finland}
\author{J. P. Pekola}
\affiliation{QTF centre of excellence, Department of Applied Physics, Aalto University School of Science, P.O. Box 13500, 00076 Aalto, Finland}
\author{A. N. Jordan}
\affiliation{Department of Physics and Astronomy, University of Rochester, Rochester, NY 14627, USA}
\affiliation{Institute  for  Quantum  Studies,  Chapman  University,  Orange,  CA  92866,  USA}

\date{\today}

\begin{abstract}
Unveiling the impact in thermodynamics of the phenomena specific to quantum mechanics is a crucial step to identify fundamental costs for quantum operations and quantum advantages in heat engines. 
We propose a two-reservoir setup to detect the quantum component in the heat flow exchanged by a coherently driven atom with its thermal environment. Tuning the driving parameters switches on and off the quantum and classical contributions to the heat flows, enabling their independent characterization. We demonstrate the feasibility of the measurement in a circuit-QED setup. Our results pave the road towards the first experimental verification of this quantum thermodynamic signature ubiquitous in quantum technologies.
\end{abstract}

\maketitle

The burst of quantum technologies has induced a growing interest in investigating quantum properties from a thermodynamic standpoint. Particular attention has been devoted to characterize the signatures of phenomena such as quantum coherence and entanglement in thermodynamic behavior of quantum systems. Motivations include the search for quantum advantages in heat engines \cite{Klatzow19}, and the need to understand better the fundamental costs for quantum protocols in presence of an environment. 
Focusing on the impact of coherent superpositions, studies have demonstrated how coherences built up during a unitary transformation in the instantaneous energy eigenbasis can lead to additional heat dissipation during a subsequent thermalization step (inner friction). This mechanism was shown to degrade the performances of Otto engines \cite{Kosloff02,Alecce15,Thomas14} and refrigerators \cite{Pekola19}.
During a realistic implementation of an engine stroke involving coherent driving of a system, or of a computation gate, the driving and dissipation occur simultaneously. In this case, off-diagonal density matrix elements in the free-system eigenbasis are continuously built up by the drive and erased by bath. Such mechanism was recently shown to lead to a quantum component in the dissipated heat flow even in the simplest case of signle qubit Rabi oscillations \cite{Elouard20}. Despite coherent manipulation of a qubit in presence of an environment is at the core of most realistic quantum protocols, this effect has not been observed so far. The challenge is to separate this quantum contribution from the classical heat flow, often of larger magnitude. 

Recent progress in the field of heatronics, i.e. the management of heat flow at the nanoscale, give opportunities to pass this bottleneck. Last advances include the design and experimental realization of nanoscale thermal rectifiers \cite{Scheibner08,Ruokola11,Fornieri14,Jiang15,Sanchez15,Martinez-Perez15}, thermal transistors \cite{Li06,Joulain16,Sanchez17,Zhang18,Guo18,Tang19,Yang19} and nanoelectronic heat engines \cite{Sanchez11,Sothmann14,Koski15,Whitney18,Erdman18,Manikandan19,Haack19}. Even more recently, nanoscale heat manipulation was associated with the properties of quantum circuits and fine temperature measurements \cite{Maillet20,Ronzani18}. The emergence of such hybrid platform promises novel tools for nanoscale heat manipulation such as quantum heat switches \cite{Karimi17,Farsani19,Sothmann17}, where tuning an external parameter (namely a magnetic flux applied to transmon qubits) allows one to dramatically change the value of the heat transfer that flows through the qubits when they are coupled to heat baths (resistors). 
In this letter, we show that a setup analogous to such quantum heat switches provides a path to measure the quantum signatures in the heat flow dissipated by a single coherently-driven qubit in a thermal reservoir. In our realistic setup, changing the parameters of the external driving (intensity and frequency) switches on and off the classical and quantum contributions to the heat flow, giving the unique opportunity to characterize them independently. This effect is obtained by coupling the driven to two photonic thermal reservoirs at different temperatures. The classical contribution to the heat flow exchanged with the hot reservoir can be switched off by adjusting the intensity of the driving to stabilize the same qubit population as at thermal equilibrium. Conversely, the quantum contribution associated with the presence of coherences in the eigenbasis of the thermal equilibrium state can be independently varied by changing the detuning between the driving and qubit frequencies. We demonstrate the feasibility of the device and of the measurements of both contributions to the heat flow by analyzing a precise implementation, based on a superconducting charge qubit coupled to two resistors measured via NIS thermometry.

\textit{Setup}.-- We consider a two-level quantum system (hereafter called qubit) of frequency $\omega_0$ weakly coupled to two thermal baths ${\cal R}_\text{h}$ and ${\cal R}_\text{c}$ of temperatures $T_\text{h}>T_\text{c}$. The dynamics of the qubit is governed by the Lindblad master equation 
\bb
\dot \rho = -i[H_0,\rho]+{\cal L}_\text{h}[\rho]+{\cal L}_\text{c}[\rho]\label{eq:ME0},
\ee
where $H_0 = \hbar\omega_0\sigma_z/2$ is the Hamiltonian of the qubit and
\bb
{\cal L}_{\text{h,c}}[\rho] = \gamma_{\text{h,c}}(\bar n_{\text{h,c}}+1){\cal D}_{\sigma_-}[\rho] + \gamma_{\text{h,c}}\bar n_{\text{h,c}}{\cal D}_{\sigma_+}[\rho],\label{Lhc}
\ee
with ${\cal D}_X[\rho] = X\rho X^\dagger - \tfrac{1}{2}\{X X^\dagger,\rho\}$ the dissipation superoperator. At steady state, a heat current flows from the hot bath to the qubit, given by \cite{Alicki79} $J^\infty_{\text{h},0} = \text{Tr}\{{\cal L}_\text{h}[\pi_0]H_0\} = -(\gamma_\text{h}\hbar\omega_0/2)(z_0+1/(2n_\text{h}+1))$,
 where $\pi_0 = \frac{\mathbb{I}}{2}+z_0 \frac{\sigma_z}{2}$ 
 is the steady state of the qubit master equation, characterized by $z_0 = - (\gamma_\text{h}+\gamma_\text{c})/[\gamma_\text{h}(2\bar n_\text{h} + 1)+\gamma_\text{c}(2\bar n_\text{c} + 1)]$. So far, the coherences in the basis $\{\ket{e},\ket{g}\}$ -- which is the eigenbasis of $H_0$, but also of the thermal equilibrium states $\pi^\text{eq}_\text{h,c} = e^{-H_0/k_BT_\text{h,c}}/Z_\text{h,c}$ with the hot and cold reservoir -- play no role in the thermodynamics.


 \begin{figure}[t]
\begin{center}
\includegraphics[width=0.47\textwidth]{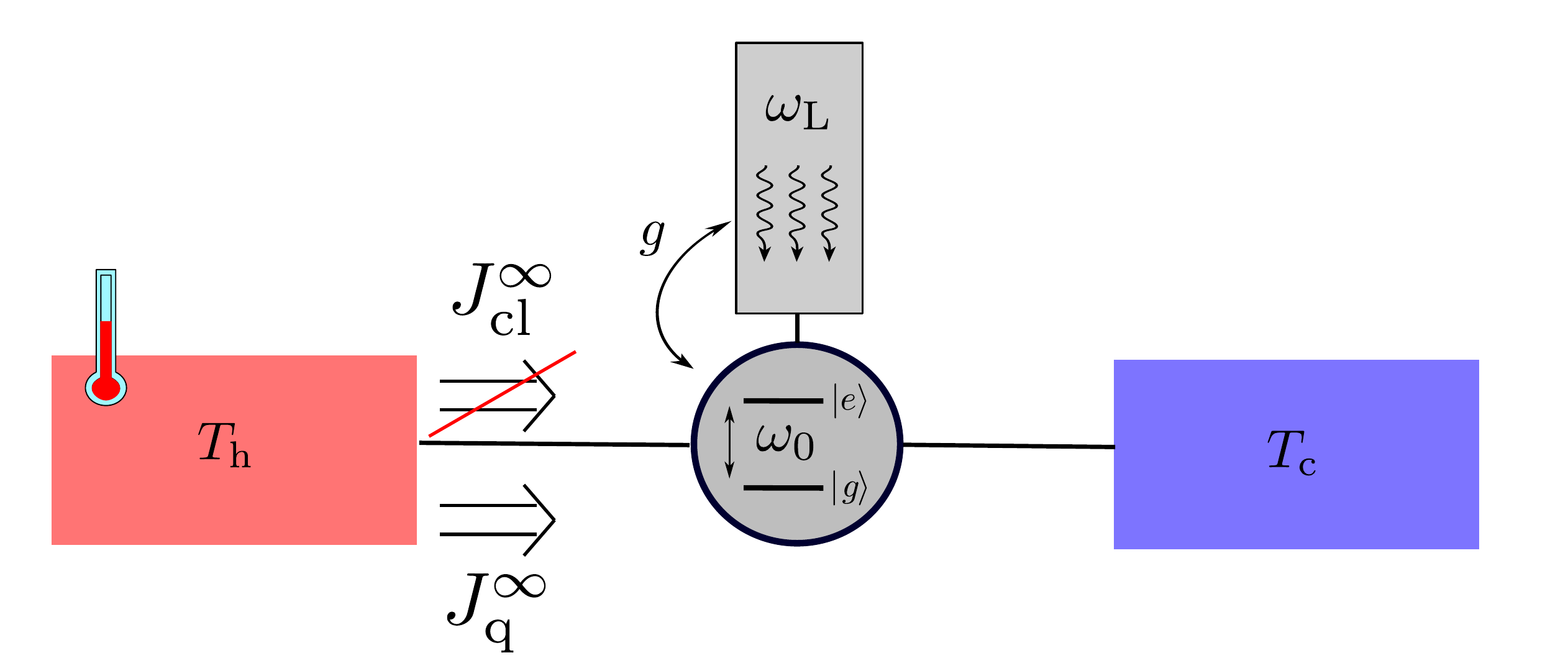}
\end{center}
\caption{Setup under investigation. The qubit is driven coherently quasi-resonantly at frequency $\omega_\text{d} = \omega_0 - \delta$, the driving-qubit coupling strength being denoted $g$, and coupled to two thermal reservoirs at temperatures $T_\text{h} > T_\text{c}$. The heat flow exchanged with the hot bath can be decomposed into a classical $J_\text{cl}$ and a quantum-coherent $J_\text{q}$ component, the latter being related to environment-induced irreversible coherence erasure. When choosing the drive to fulfill $g =g^*(\delta)$ (see text), with $\delta$ the detuning, the classical component $J_\text{cl}$ can be completely suppressed and the quantum component can be measured by thermometry of the hot bath.
\label{f:Setup}
}
\end{figure}

 We now suppose that the qubit is quasi-resonantly driven by a monochromatic field of frequency $\omega_\text{d}$, which can be modelled by adding a time-dependent term $H_\text{d}(t) = \dfrac{\hbar g}{2}\left(e^{i\omega_\text{d}t}\sigma_- + e^{-i\omega_\text{d}t} \sigma_+\right)$ in its Hamiltonian. We have denoted $g$ the field-matter coupling strength. This term induces a rotation of the state of the qubit in the Bloch sphere along the rotating unit vector $\vec u(t) = (\cos(\omega_\text{d}t),\sin(\omega_\text{d} t),0)$. In the limit where $g,\vert\delta\vert \ll \omega_0, \omega_\text{d}$, with $\delta = \omega_0-\omega_\text{d}$ the detuning, and provided the spectral densities of the reservoirs are flat around the frequency $\omega_0$, the dissipation induced by the bath is unchanged by the presence of the drive (see \cite{CCT,Elouard20}). Therefore, the evolution of the density operator of the qubit is ruled by the same master equation as before (see Eq.~\eqref{eq:ME0}) except that the Hamiltonian part of the dynamics is generated by $H_0+H_\text{d}(t)$ instead of $H_0$. 
 
 Interestingly, $H_\text{d}(t)$ continuously generates coherences in the thermal equilibrium eigenbasis $\{\ket{e},\ket{g}\}$, while the action of the two baths is to continuously erase them. The competition between the driving and the dissipation results in a stationary orbit with non-zero coherences. It takes the form $\pi(t) = U_\text{rot}^\dagger\tilde \pi U_\text{rot}$, where $U_\text{rot} = e^{it\omega_\text{d}\sigma_z/2}$ is the unitary transformation to the frame rotating at the driving frequency and $\tilde \pi =  (\mathbb{I}+\vec r_\infty\cdot \vec \sigma)/2$ is the steady state reached by the qubit in such rotating frame. We have denoted $\vec r_\infty = (\tilde x_\infty,\tilde y_\infty,\tilde z_\infty)$ the steady-state Bloch vector in the rotating frame and $\vec\sigma = (\sigma_x,\sigma_y,\sigma_z)$ the vector of Pauli matrices. The exact expression of the steady state can be found analytically, yielding:
 \begin{subequations}
\bb
\tilde x_\infty &=& - \frac{2 \delta g (\gamma_\text{h}+\gamma_\text{c})/\gamma_\text{tot}}{2 g^2 + \gamma_\text{tot}^2 + 4\delta^2},\label{xinf}\\
\tilde y_\infty &=&  \frac{g (\gamma_\text{h}+\gamma_\text{c})}{2 g^2 + \gamma_\text{tot}^2 + 4\delta^2},\label{yinf}\\
\tilde z_\infty &=&  -\frac{(\gamma_\text{h}+\gamma_\text{c})(\gamma_\text{tot}^2 + 4\delta^2)}{\gamma_\text{tot}(2 g^2 + \gamma_\text{tot}^2 + 4\delta^2)},\label{zinf}
\ee
\end{subequations}
with $\gamma_\text{tot} = \gamma_\text{h}(2n_\text{h}+1)+\gamma_\text{c}(2n_\text{c}+1)$.
 
 Note that in contrast with $\pi_0$, the stationary orbit state carries non-zero average value of the coherences of constant modulus $\vert\bra{e}\pi(t)\ket{g}\vert = \sqrt{\tilde x_\infty^2+ \tilde y_\infty^2}$ in the free-qubit energy eigenbasis $\{\ket{e},\ket{g}\}$, where $\sigma_z =\ket{e} \bra{e}-\ket{g}\bra{g}$. These coherences are characterized by a contribution in phase with the driving $\tilde x_\infty= \text{Tr}\{(\vec u(t).\vec\sigma)\pi(t)\}$ and out of phase $\tilde y_\infty = \text{Tr}\{(\vec v(t)\cdot\vec\sigma)\pi(t)\}$, with $\vec v(t) = (-\sin(\omega_\text{d}t),\cos(\omega_\text{d}t),0)$ is a vector orthogonal to $\vec u(t)$.

 \textit{Quantum contribution to the heat flow}.-- In presence of the quasi-resonant drive, the heat flow, defined as the energy provided by the hot reservoir to the qubit, takes the value \cite{Elouard20} $J_\text{h}(t) =  J_\text{cl}(t) + J_\text{q}(t) = \text{Tr}\{H_0{\cal L}_\text{h}[\rho(t)]\} + \text{Tr}\{H_\text{d}(t){\cal L}_\text{h}[\rho(t)]\}$. The contribution $J_\text{cl}(t) = -(\gamma_{\text{h}}\hbar\omega_0/2)(z(t)+ 1/(2\bar n_\text{h}+1))$ is similar to the undriven case and can be interpreted as the heat flow in the case of a classical two-level system, unable to carry coherences in the $\{\ket{e},\ket{g}\}$ basis. Conversely, the contribution $J_\text{q}(t) = - \gamma_\text{h}(2\bar n_\text{h}+1)\hbar g \tilde x(t)/4$, with $\tilde x(t) = \text{Tr}\{(\vec u(t)\cdot\vec \sigma)\rho(t)\}$, is proportional to the amplitude of the coherences in the $\{\ket{e},\ket{g}\}$ basis and is therefore a genuinely quantum contribution. It encompasses the price for the reservoir to erase the coherences in phase with the driving. The latter indeed contribute to the energy stored in the qubit's state via the term $E_\text{q}(t) = \text{Tr}\{H_\text{d}(t)\rho(t)\} = \hbar g \tilde x(t)/2$. The coherences out of phase $\tilde y(t)= \text{Tr}\{(\vec v(t)\cdot\vec \sigma)\rho(t)\}$ do not contribute to the qubit's energy and do not play any role in the heat flow. A more detailed analysis of this contribution and the thermodynamics of the driven qubit can be found in Ref.~\cite{Elouard20}. For now, let us note that the classical component scales like $\omega_0$, while the quantum contribution scales like the Rabi frequency $g$ which is necessary of much smaller magnitude in the situation we are interested in (and for which the present dynamical model is valid). This scaling suggests that it is in general very difficult to separate both contributions from each other and observe quantum effects in the heat flow from a driven quit.
 
 \begin{figure}
\begin{flushleft}
\includegraphics[width=0.5\textwidth]{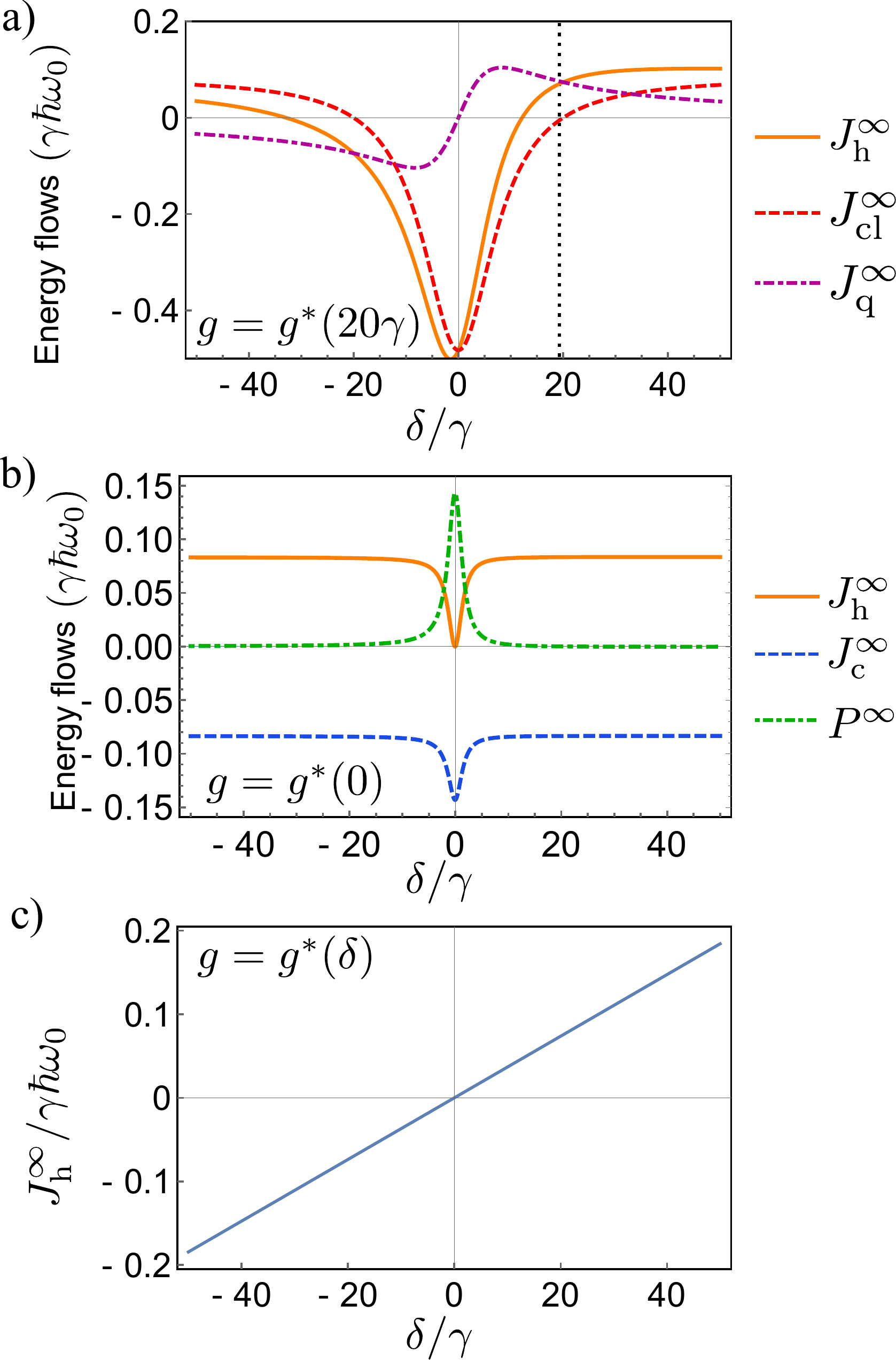}
\end{flushleft}
\caption{\textbf{a}: Classical $J_\text{cl}^\infty$ and quantum $J_\text{q}^\infty$ contributions to the stationary heat flow $J_\text{h}^\infty$ (solid orange) provided by the hot thermal reservoir for $g = g^*(20\gamma)$, as a function of the detuning $\delta$. The dotted vertical line indicates the value of the detuning $\delta = 20\gamma$ where the classical heat vanishes. \textbf{b}: Stationary heat flow $J_\text{h}^\infty$ (solid orange) provided by the hot thermal reservoir, $J_\text{h}^\infty$ (dashed blue) provided by the cold thermal reservoir and power $P^\infty$ injected by the drive (dot-dashed green) for $g = g^*(0)$ (see Eq.~\eqref{eq:gst}) as a function of the detuning $\delta$.\textbf{c}: Total heat flow from the hot bath $J^\infty_\text{h}$ along the line $(g^*(\delta),\delta)$ of parameters. Parameters: $\gamma_\text{h} = \gamma_\text{c} = \gamma =2.7$~GHz, $\omega_0 = 2\pi\times 10$~GHz, $\bar n_\text{h} = 0.34$, $\bar n_\text{c} =0.10$.
}\label{f:EnergyPlots}
\end{figure} 
 
\textit{Switching independently the quantum and classical heat flow}.-- The stationary value of the heat flow can be controlled by engineering the steady-state of the qubit, which in turn can be adjusted by tuning the driving parameters, namely the coupling strength $g$ (determined by the driving intensity) and the detuning $\delta$. We first show that the classical part of the heat flow can be completely switched off. The key idea is that this contribution is zero if the population of the qubit in the $\{\ket{e},\ket{g}\}$ basis matches the thermal equilibrium with the hot reservoir, i.e. $z^\text{eq}_\text{h} = -1/(2\bar n_\text{h}+1)$. For each fixed value of the detuning $\delta$, this can be realized for a particular value of the driven intensity $g^*(\delta)$ found by solving the equation $\tilde z_\infty = z^\text{eq}_\text{h}$, yielding:
 \bb
g^*(\delta) = \left[\frac{\gamma_\text{c}}{\gamma_\text{tot}}(\gamma_\text{tot}^2+4\delta^2)(\bar n_\text{h}-\bar n_\text{c})\right]^{1/2}.\label{eq:gst}
\ee
From the proportionality to the square root of the thermal occupation difference $\bar n_\text{h}-\bar n_\text{c}$, it is clear that the classical part of the heat current can be suppressed solely in the presence of a colder bath at temperature $T_\text{c}< T_\text{h}$.
Even when the population of the qubit matches its value at thermal equilibrium with the hot bath, the stationary state still differs from the thermal equilibrium state $\pi^\text{eq}_\text{h}$ because of coherences in the $\{\ket{e},\ket{g}\}$ basis. This results in a non-zero value of the stationary quantum contribution $J_\text{q}^\infty$ of the heat current: the present setup therefore allows us to separate the classical and quantum contribution by canceling $J_\text{cl}^\infty$. Setting $g=g^*(\delta)$ and measuring the slight temperature variations of the hot reservoir provides a method to measure the quantum contribution to the heat flow (see also experimental proposal below). The steady state quantum heat flow takes the value:
\bb
J_\text{q}^\infty = \hbar\delta\frac{\gamma_\text{h}\gamma_\text{c}}{\gamma_\text{tot}}(\bar n_\text{h}-\bar n_\text{c}).\label{Jq}
\ee
We stress that separating the contributions is impossible when there is only one thermal bath (or equivalently when $T_\text{c}=T_\text{h}$). The latter situation implies $\vert J_\text{q}^\infty\vert \ll \vert J_\text{cl}^\infty\vert$ for any choice of parameters, except for $g=0$ where both contributions vanish. The two contributions of the heat flow $J_\text{h}^\infty$ are plotted in Fig.~\ref{f:EnergyPlots}.\textbf{a} as a function of the detuning $\delta$ for $g=g^*(\gamma)$. We stress that in contrast to the inner friction which always corresponds to heat dissipated in the environment \cite{Plastina14}, the quantum component of the heat can be either positive or negative and can therefore increase or decrease the steady state entropy production $\dot\sigma_\infty = -(J_\text{q}^\infty+ J_\text{cl}^\infty)/T$, respectively. At steady state, it takes the sign of the detuning (see Eq.~\eqref{Jq}).

Due to its proportionality to the in-phase coherences $\tilde x_\infty$, the quantum contribution vanishes when the qubit is driven exactly at resonance $\delta =0$ (see Eq.~\eqref{xinf}), so that for $g>0$ and $\delta =0$, the heat flow only contains the classical contribution. A special point corresponds to $(g,\delta) = (g^*(0),0)$ where both contributions  to $J_\text{h}$ are zero even though the qubit state differs from the thermal equilibrium state at temperature $T_\text{h}$. Indeed, the out-of-phase coherences equal for these parameters $\tilde y_\infty = ((\bar n_\text{h}-\bar n_\text{c})\gamma_\text{c}/\gamma_\text{tot})^{1/2}/(2\bar n_\text{h}+1)$. Changing the driving parameters slightly around this special point allow to switch off and on the total heat flow from the hot bath, the device effectively acting in a way similar to a heat switch. We stress however that the device requires constant power input: The power provided by the driving $P^\infty = \text{Tr}\{\dot H_\text{d}(t)\pi(t)\}$ takes the value $\hbar\omega_0\gamma_\text{c}(\bar n_\text{h}-\bar n_\text{c})/(2\bar n_\text{h}+1)$ at the special point. This value is positive, meaning that a constant amount of power is provided to the qubit and eventually dissipated in the cold bath. The device can therefore stop heat to flow from the hot bath to the qubit, but does not protect the cold bath from receiving heat. This effect is illustrated in Fig.~\ref{f:EnergyPlots}.\textbf{b} where the two component of the heat flow from the hot bath and the power injected by the drive are plotted against the detuning for $g = g^*(0)$.


\textit{Characterization of the quantum heat.} -- When setting $g = g^*(\delta)$, the quantum heat contribution to the heat can be measured by monitoring the slow temperature variations of the hot bath. This contribution can be distinguished from a residual classical heat flow owing to its linear dependence on the detuning $\delta$ when following the line $(g^*(\delta),\delta)$ (see Fig.~\ref{f:EnergyPlots}.\textbf{c}), whereas the classical heat follows the Lorentzian dependence of $z^\infty$.  

\begin{figure}
\begin{center}
\includegraphics[width=0.38\textwidth]{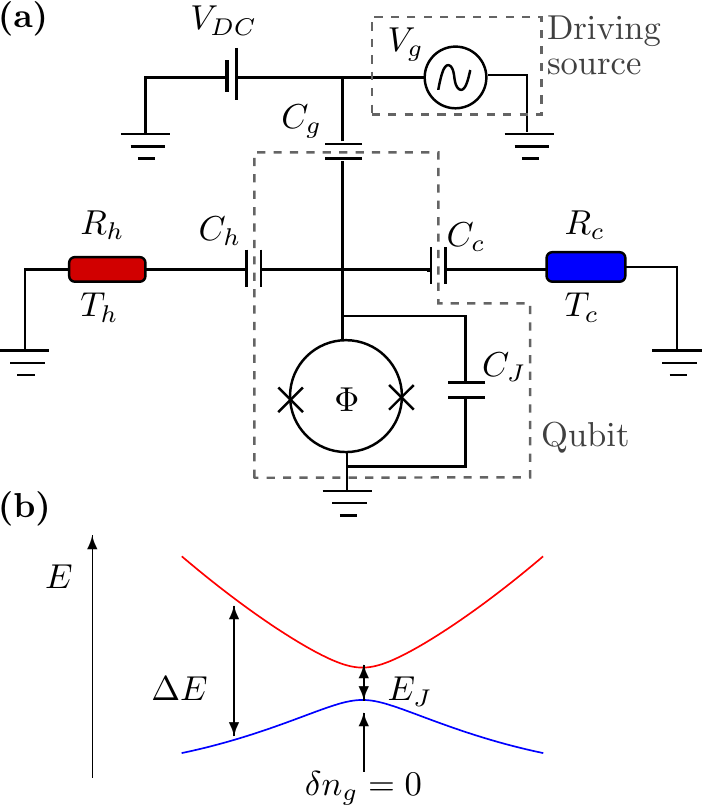}
\end{center}
\caption{
(a) Schematic experimental set up for the implementation of a quantum heat switch in a superconducting circuit: A driven SQUID loop is capacitively coupled to a hot and cold resistors.
(b) First and second energy levels (for $E_c>E_J$)  versus $ \delta n_g$ are plotted. We only drive the system very close to $\delta n_g=0$.}
\label{Exp_dgm}
\end{figure}

\textit{Implementation in a superconducting circuit.} -- We now analyze the feasibility of the scheme in a typical superconducting quantum circuit setup.
Superconducting qubits are versatile candidates to perform different quantum thermodynamic experiments
as they can be  controlled externally  and measured with high precision \cite{Koch2007,Alberto2018,Thomas2019}.
 The superconducting qubit (two-level system) can be a transmon \cite{Koch2007}, a flux qubit \cite{Mooij1999} or a charge qubit \cite{Wallraff2004} based on a Cooper pair box \cite{Buttiker87,Bouchiat1998}. In this Letter, we discuss the implementation of the heat switch using a charge qubit, which is based on a superconducting island with very small capacitance. This island is capacitively coupled to the baths and driving electrode, and terminated by a superconducting loop made of two small Josephson junctions, as shown in Fig.~\ref{Exp_dgm}. Once quantized, the circuit behaves as an anharmonic oscillator such that the two lower levels can be addressed independently from other levels and treated as a qubit. For an appropriate choice of constant voltage $V_{DC}$, the Hamiltonian of the qubit in the charge basis is given as
\begin{equation}
H= E_C \delta n_g(t) \sigma_z -\frac{E_J}{2} \sigma_x.
\end{equation}
where the charging energy $E_C=e^2/2C_\Sigma$, the total capacitance of the island $C_\Sigma=C_c+C_h+C_g+C_J$, and $E_J=E_J(\Phi)$ is the Josephson energy that can be tuned with an external flux $\Phi$. 
Note that with respect to the general analysis above, the free qubit quantization axis and the driving axis have been swapped to match usual conventions in the field. Here $\delta n_g (t)=C_gV_g/(2 e)$ is driven near the avoided level crossing $\delta n_g= 0$,
with a  voltage  $V_g \cos{(\omega_L t)}$  and gate capacitance is $C_g$.
At $\delta n_g= 0$, the energy gap is $E_J$. Further, two normal-metal resistors acts as the heat baths whose temperatures can be controlled and measured
with Normal-Insulator-Superconductor (NIS) tunnel junctions which are sensitive local electronic heaters and thermometers \cite{Giazotto06}. 
The qubit is capacitively coupled to the heat baths in order to achieve a weak coupling of the qubit with the environment.
The transition rates are given by $\gamma_\text{h,c}\approx 2\pi C_\text{h,c}^2 E_J R_\text{h,c}/(2 \hbar C_{\Sigma}^2 R_Q)$, where $R_Q=h/4 e^2$ is the superconducting resistance quantum~\cite{Bouchiat1998,Lehnert03,Pekola_Golubev2016}. 
The temperatures of the baths can be taken in the range of 30 mK to 350 mK such that the populations in the higher excited states of the qubit can be ignored.
Realistic driving parameters (see Supplement \cite{SI}) $(\delta/2 \pi,g^*(\delta)/2 \pi) \sim (0.1,0.4)~\mathrm{GHz}$  correspond to quantum heat contribution releasing $J_\text{q}\sim \numprint{1.5E-17}~\mathrm{W}$ in the resistor.
In this setup, there is a background heat flow from the  hot bath  to the cold bath through the capacitors and phonons which is ignored in this context.

\textit{Discussion and conclusion}: The scheme presented is still valid in presence of pure dephasing at rate $\gamma_{\phi}$. In this case, the value of the driving strength $g^*(\delta)$ allowing to suppress the classical heat flow has an expression similar to Eq.~\eqref{eq:gst}, except replacing $\gamma_\text{tot}$ with $\gamma_\text{tot}+2\gamma_\phi$ (see Supplement \cite{SI}). The quantum heat flow is still canceled at resonance. In fact, the charge qubit setup may allow a good test of this property, since detuning the qubit away from the avoided level crossing  makes the qubit much more sensitive to charge noise and therefore increases the dephasing rate \cite{Ithier2005}. 

In this Letter, we have demonstrated that a two-temperature setup allows to measure and characterize independently the quantum and classical contributions in the heat flow dissipated by a driven qubit. We have proposed an implementation of the scheme in a circuit QED setup and analyzed its feasibility in a state-of-the-art setup. Such an observation opens the path to a better understanding of the thermodynamic costs of quantum operations and experimental comparisons of the performances of quantum and classical heat engines in the coherent regime of operations.

{\textit{Acknowledgements:} We thank Dmitry S. Golubev and Bayan Karimi for discussions. Work by CE and ANJ was supported by the US Department of Energy (DOE), Office of Science, Basic Energy Sciences (BES), under Grant No. DE-SC0017890. G. T., O. M. and J. P. P. acknowledge the support from the Academy of Finland Centre of Excellence
program (project 312057) and the European Union's Horizon 2020 research and innovation programme under the European Research Council (ERC) programme (grant agreement 742559).

\pagebreak

\onecolumngrid

\section*{Quantifying the quantum heat contribution from a driven superconducting circuit: Supplementary material}

\subsection*{Steady state of the qubit evolution}

The master equation for the atom Eq.~(3) of main text, when written in the rotating frame, can be expressed in terms of the vector $Y(t) = (P_e(t),\tilde x(t),\tilde y(t),P_g(t))^\top$, with $P_e(t) = (z(t)+1)/2$ and $P_g(t) = 1-P_e(t)$, leading to:
\bb
\dot Y(t) = A Y(t) \label{eq:LODE}
\ee
where

\bb
A = \left(
\begin{array}{cccc}
-\gamma_\text{h}(n_\text{h}+1)-\gamma_\text{c}(n_\text{c}+1) & 0 & g & \gamma_\text{h}n_\text{h}+\gamma_\text{c}n_\text{c} \\ 
0 & -\frac{\gamma_\text{h}}{2}(2n_\text{h}+1)-\frac{\gamma_\text{c}}{2}(2n_\text{c}+1) & -\delta & 0 \\ 
-\frac{g}{2} & \delta & -\frac{\gamma_\text{h}}{2}(2n_\text{h}+1)-\frac{\gamma_\text{c}}{2}(2n_\text{c}+1) & \frac{g}{2} \\ 
\gamma_\text{h}(n_\text{h}+1)+\gamma_\text{c}(n_\text{c}+1) & 0 & -g & -\gamma_\text{h}n_\text{h}-\gamma_\text{c}n_\text{c}
\end{array}\right)\quad\quad. 
\ee

Solving Eq.~\eqref{eq:LODE} for its stationary state allows to derive Eqs.~(4a)-(4c).

\subsection*{Effect of additional decoherence channel}

A dephasing channel causing a decay of the coherences in the $\{\ket{e},\ket{g}\}$ basis of the qubit at a rate $\gamma_\phi$ can be modeled by an additional term ${\cal L}_\phi[\rho] = (\gamma_\phi/2) D_{\sigma_z}[\rho]$ in the master equation of the qubit Eq.~(3). This results in a new evolution for $Y(t)$, given by $\dot Y(t) = A_1 Y(t)$, with 

\bb
A_1= \left(
\begin{array}{cccc}
-\gamma_\text{h}(n_\text{h}+1)-\gamma_\text{c}(n_\text{c}+1) & 0 & g & \gamma_\text{h}n_\text{h}+\gamma_\text{c}n_\text{c}-\gamma_\phi \\ 
0 & -\frac{\gamma_\text{h}}{2}(2n_\text{h}+1)-\frac{\gamma_\text{c}}{2}(2n_\text{c}+1)-\gamma_\phi & -\delta & 0 \\ 
-\frac{g}{2} & \delta & -\frac{\gamma_\text{h}}{2}(2n_\text{h}+1)-\frac{\gamma_\text{c}}{2}(2n_\text{c}+1) & \frac{g}{2} \\ 
\gamma_\text{h}(n_\text{h}+1)+\gamma_\text{c}(n_\text{c}+1) & 0 & -g & -\gamma_\text{h}n_\text{h}-\gamma_\text{c}n_\text{c}
\end{array}\right)\quad\quad. 
\ee
The new steady state of the qubit (in the rotating frame) is then:
\bb
\tilde x_\infty &=& - \frac{4 \delta g (\gamma_\text{h}+\gamma_\text{c})}{2 g^2(\gamma_\text{tot}+2\gamma_\phi)+ \gamma_\text{tot}((\gamma_\text{tot}+2\gamma_\phi)^2 + 4\delta^2)}\label{xinfphi}\\
\tilde y_\infty &=&  \frac{2g (\gamma_\text{h}+\gamma_\text{c})(\gamma_\text{tot}+2\gamma_\phi)}{2 g^2(\gamma_\text{tot}+2\gamma_\phi) + \gamma_\text{tot}((\gamma_\text{tot}+2\gamma_\phi)^2 + 4\delta^2)}\label{yinfphi}\\
\tilde z_\infty &=&  -\frac{(\gamma_\text{h}+\gamma_\text{c})((\gamma_\text{tot}+2\gamma_\phi)^2 + 4\delta^2)}{2 g^2(\gamma_\text{tot}+2\gamma_\phi) + \gamma_\text{tot}((\gamma_\text{tot}+2\gamma_\phi)^2 + 4\delta^2)}\label{zinfphi},
\ee

As before, we solve $(\tilde z_\infty+1)/2 = P_e^\text{h}$ to find the value of $g$ canceling the classical heat flow from the hot bath. We obtain:
 \bb
g^*(\delta) = \left[\frac{\gamma_\text{c}}{\gamma_\text{tot}+2\gamma_\phi}\Big((\gamma_\text{tot}+2\gamma_\phi)^2+4\delta^2\Big)(\bar n_\text{h}-\bar n_\text{c})\right]^{1/2}.\label{eq:gstphi}
\ee

The heat flow from the hot bath $J_\text{h}^\infty = \text{Tr}\{H{\cal L}_\text{h}[\rho]\}$ then reads for $g=g^*(\delta)$:
\bb
J_\text{h}^\infty = J_\text{q}^\infty = \hbar\delta\frac{\gamma_\text{h}\gamma_\text{c}}{\gamma_\text{tot}+2\gamma_\phi}(\bar n_\text{h}-\bar n_\text{c}),
\ee
such that as for $\gamma_\phi = 0$, the choice $\delta = 0$ allows to completely stop the heat current from the hot bath.

%

\subsection*{Feasible experimental parameters}
\label{Exp_par}

The capacitances $C_h$ and $C_c$ must be chosen so that they do not exceed $C_J$, the Josephson junctions' capacitance (because the charging energy should be high enough), while  allowing reasonable coupling to resistors $R_h$ and $R_c$. Meanwhile, $C_g$ should satisfy $C_g \ll C_J$. The charging energy is $E_C= e^2/2C_{\Sigma}$. A  choice of experimentally accessible parameters that may satisfy all these requirements is: $C_J = 2 \times 0.3$ fF (factor 2 because of two junctions), $C_h=C_c = 0.3$ fF, $C_g=0.03$ fF. For the resistors, we can use a large range depending on the desired relaxation rate. Highly sensitive measurements needs high resistances. 
For example, AuPd resistors, which enable NIS thermometry, allows to reach the resistance value around $1.5~\text{k}\Omega$. With a set of parameters chosen so as to maximize $\gamma_\text{h,c}$, and supposing a typical Josephson energy $E_J/h=10$ GHz, one obtains $\gamma_\text{h,c}\approx 2.7$ GHz, which can be easily varied with the resistances values. Considering bath temperatures, $T_h=350$ mK, $T_c=200$ mK, and detuning $\delta/2 \pi=0.1$ GHz, the amplitude of the drive is $g^*/(2 \pi) \approx 0.4$ GHz which leads to $J_q\approx 15$ aW.

\end{document}